\documentclass[letter]{aa} 
\newcommand{\nc}{\newcommand}
\nc{\lra}{\leftrightarrow}
\nc{\cm}{\,\mathrm{cm}}
\nc{\micron}{\,\mathrm{\mu m}}
\nc{\cmcm}{\,\mathrm{cm^{-2}}}
\nc{\cmcmg}{\,\mathrm{cm^{2}\,g^{-1}}}
\nc{\iccm}{\,\mathrm{cm^{-3}}}
\nc{\cul}{\,\mathrm{cm^{3}\, s^{-1}}}
\nc{\kmps}{\,\mathrm{km\, s^{-1}}}       
\nc{\Kkmps}{\,\mathrm{K\, km\, s^{-1}}}       
\nc{\K}{\,\mathrm{K}}       
\nc{\vlsr}{v_\mathrm{LSR}}       
\nc{\etamb}{\eta_\mathrm{mb}}       
\nc{\vt}{v_\mathrm{turb}}       
\nc{\dv}{\Delta v}       
\nc{\trot}{T_\mathrm{rot}}      
\nc{\tkin}{T_\mathrm{kin}}      
\nc{\tex}{T_\mathrm{ex}}      
\nc{\nhh}{n(\mathrm{H_2})}      
\nc{\Nhh}{N(\mathrm{H_2})}      
\nc{\nmol}[1]{N_\mathrm{#1}}      
\nc{\Xmol}[1]{$X[\mathrm{#1}]$} 
\nc{\ebf}{\eta_\mathrm{bf}}      
\nc{\jkk}[4]{${#1}_{#2}-{#3}_{#4}$}
\nc{\me}[2]{#1\times 10^{#2}}
\nc{\hh}{$\mathrm{H_2}$}
\nc{\hooh}{$\mathrm{HOOH}$}
\nc{\hho}{$\mathrm{H_2O}$}
\nc{\hoo}{$\mathrm{HO_2}$}
\nc{\hhco}{$\mathrm{H_2CO}$}
\nc{\hhico}{$\mathrm{H_2^{13}CO}$}
\nc{\hdco}{$\mathrm{HDCO}$}
\nc{\ddco}{$\mathrm{D_2CO}$}
\nc{\chhhoh}{$\mathrm{CH_3OH}$}
\nc{\nhhh}{$\mathrm{NH_3}$}
\nc{\oo}{$\mathrm{O_2}$}
\nc{\oio}{$\mathrm{O^{18}O}$}
\nc{\co}{$\mathrm{CO}$}
\nc{\cio}{$\mathrm{C^{18}O}$}
\nc{\icio}{$\mathrm{^{13}C^{18}O}$}
\nc{\ico}{$\mathrm{^{13}CO}$}
\nc{\so}{$\mathrm{SO}$}
\nc{\iso}{$\mathrm{^{34}SO}$}
\nc{\isoo}{$\mathrm{^{34}SO_2}$}
\nc{\soo}{$\mathrm{SO_2}$}
\nc{\nndp}{$\mathrm{N_2D^+}$}
\nc{\nnhp}{$\mathrm{N_2H^+}$}
\nc{\ccchh}{$c-\mathrm{C_3H_2}$}
\nc{\rhooph}{$\rho$~Ophiuchi}
\nc{\roa}{$\rho$~Oph~A}
\nc{\roc}{$\rho$~Oph~C}
\nc{\mc}[1]{\multicolumn{1}{c}{#1}}
\nc{\sref}[1]{Sect.~(\ref{#1})}
\nc{\fref}[1]{Fig.~\ref{#1}}
\nc{\frefs}[2]{Figs.~\ref{#1} and \ref{#2}}
\nc{\tref}[1]{Table~\ref{#1}}
\nc{\eref}[1]{Eq.~(\ref{#1})}
\nc{\RA}[4]{$\alpha(\mathrm{#1})={#2}^\mathrm{h}\,{#3}^\mathrm{m}\,{#4}^\mathrm{s}$}
\nc{\Dec}[4]{$\delta(\mathrm{#1})={#2}\degr\,{#3}\arcmin\,{#4}\arcsec$}
\usepackage{graphicx}
\usepackage{txfonts}
\usepackage{natbib}
\bibpunct{(}{)}{;}{a}{}{,} 
\begin{document}
   \title{Detection of interstellar hydrogen peroxide
   \thanks{Based on observations with the Atacama Pathfinder EXperiment (APEX)
   telescope. APEX is a collaboration between the Max-Planck-Institut f{\"u}r
   Radioastronomie, the European Southern Observatory, and the Onsala Space
   Observatory.}}


   \author{P. Bergman\inst{1}
        \and
           B. Parise\inst{2}
	\and
	   R. Liseau\inst{3}
	\and
	   B. Larsson\inst{4}
	\and
	   H. Olofsson\inst{1}
	\and
	   K. M. Menten\inst{2}
	\and
	   R. G{\"u}sten\inst{2}
        }

   \institute{Onsala Space Observatory, Chalmers University of Technology,
              SE-439 92 Onsala, Sweden
              \email{per.bergman@chalmers.se}
         \and
              Max Planck Institut f{\"u}r Radioastronomie, Auf dem H{\"u}gel 69,
	      53121 Bonn, Germany
	 \and
	      Department of Earth and Space Sciences, Chalmers University of Technology,
              SE-439 92 Onsala, Sweden\ 
	 \and
	      Department of Astronomy, Stockholm University, AlbaNova,
	      SE-10691 Stockholm, Sweden
             }

   \date{Received ?; accepted ?}

 
  \abstract
   {The molecular species hydrogen peroxide, HOOH, is likely to be a key ingredient in the oxygen
   and water chemistry in the interstellar medium.}
   {Our aim with this investigation is to determine how abundant {\hooh} is in the cloud core
   {\roa}.}
   {By observing several transitions of HOOH in the (sub)millimeter regime we seek to identify
   the molecule and also to determine the excitation conditions through a multilevel excitation
   analysis.}
   {We have detected three spectral lines toward the SM1 position of {\roa} at velocity-corrected
   frequencies that coincide very closely with those measured from laboratory spectroscopy of {\hooh}.
   A fourth line was detected at the $4\sigma$ level. We also found through mapping observations
   that the {\hooh} emission extends (about 0.05 pc) over the densest part of the {\roa}
   cloud core. We derive an abundance of {\hooh} relative to that of {\hh} in the SM1 core 
   of about $\me{1}{-10}$.
   }
   {To our knowledge, this is the first reported detection of HOOH
   in the interstellar medium.}


   \maketitle
%

\section{Introduction}
Hydrogen peroxide, {\hooh}, is believed to play an important role in the Earth's
atmospheric ozone and water chemistry. It is a key constituent in the gas- and
liquid-phase radical chemistry and has an oxidizing potential in the liquid phase.
Gas-phase {\hooh} has been seen in the
Martian atmosphere by ground-based observations \citep{2004Icar..168..116C,
2004Icar..170..424E}. However, recent Mars observations with the {\it Herschel}
Observatory \citep{2010A&A...521L..49H} failed to detect {\hooh} at levels below those
previously seen. The non-detection was attributed to seasonal variations.

Interestingly, {\hooh} is among the simplest molecules that show internal rotation. The
internal rotation, or torsion, manifests itself as a rotation of the two O-H bonds about
the O-O bond. This hindered internal rotation can be described with a torsion potential
in which the two minima (the most stable configurations) do not coincide with the
{\it cis} or {\it trans} alignment of the two O-H bonds\footnote{When the two O-H bonds
point in the same direction, this is referred to as the {\it cis} position, while the 180
degree opposite case is called the {\it trans} position. For {\hooh}, the {\it trans}
potential barrier height is 557 K, while the {\it cis} barrier height is almost 4000 K
\citep{1993JMoSp.159..507P}.}. The twofold barrier gives rise to a quartet of sublevels
for each torsional state. These sublevels are denoted $\tau=1, 2, 3, 4$
\citep{1965JChPh..42.1931H}. Moreover, {\hooh} is a light, slightly asymmetric prolate
rotor (belonging to the $C_{2h}^\dagger$ point group, see \citet{1984CaJPh..62.1392H}
for a discussion), with only $c$-type transitions and with a dipole moment of
1.6 D \citep{1981JMoSp..87..582C, 1996JMoSp.176..287P}. Because of this, no pure rotational
transitions occur, and only transitions corresponding to a combined ro-torsional motion
change can take place. The mm and submm spectrum of {\hooh} has been studied
rather extensively in the laboratory \citep{1981JMoSp..85..120H, 1995JMoSp.171..145P}
and is available in the JPL database \citep{1998JQSRT..60..883P}.

To our knowledge, {\hooh} has not so far been detected in the interstellar medium.
Also, very few abundance limits have been reported.
\citet{1987ApJ...315..621B} derived an upper limit for {\hooh} of $\me{4.5}{-10}$
with respect to {\hh} from their Orion spectral scan data.
\citet{1998A&A...331..749B} reported an upper limit of 5.2\,\% of solid {\hooh}
relative to {\hho} ice toward NGC~7538\, IRS9.

As in the Earth's atmosphere, {\hooh} is expected to be closely connected to the
water and molecular oxygen chemistry also for those physical conditions prevailing
in molecular clouds. In current pure gas-phase models, {\hooh} is formed by
reaction of {\hh} with {\hoo} or via the reaction involving two OH radicals. However,
these reactions proceed very slowly\footnote{{\tt
http://www.physics.ohio-state.edu/$\sim$eric}}. Alternatively,
\citet{1982A&A...114..245T} suggested that {\hooh} could be formed on grain
surfaces by the successive additions of H atoms to molecular oxygen. If this is
the case, {\hooh} could be closely related to the amount of {\oo} on grains.

The 119 GHz line of {\oo} was detected toward the cloud {\roa} with an abundance of
$\me{5}{-8}$ relative to {\hh} by \citet{2007A&A...466..999L} using the Odin satellite.
The {\roa} molecular cloud, at a distance of about 120 pc, has been the subject of
several studies. Continuum observations \citep{1993ApJ...406..122A, 1998A&A...336..150M} and
{\cio} observations \citep{2010A&A...510A..98L} revealed several cores. Very recently,
Bergman and coworkers found a very high degree of deuteration toward the SM1 core in
{\roa} from observations of deuterated {\hhco} \citep{2011A&A...527A..39B}. These 
observations suggest that grain surface reactions were at work to produce the very
high deuterium levels observed in the gas-phase material.

In this {\it Letter} we continue our study of the {\roa} cloud by reporting on
observations of several {\hooh} transitions. We have detected four {\hooh} lines and
two of these lines were mapped over the central part of this cloud. In
\sref{s:obsres} we describe our observations and present our results. There, we
also describe in more detail the energy level structure and symmetry of {\hooh},
which is important for discussing the detected lines. Then, in \sref{s:dis}, we discuss
the implications of our results.


\section{Observations and results}
\label{s:obsres}
We have used the APEX 12 m telescope located at about 5100\,m altitude in the
Chilean Andes \citep{2006A&A...454L..13G} to observe {\hooh}. For the lower
frequency lines we used the Swedish heterodyne facility instruments APEX-1
and APEX-2 \citep{2008A&A...490.1157V}. The 7-pixel longer wavelength ($450\micron$)
module of the MPIfR-built CHAMP$^+$ receiver array \citep{2006SPIE.6275E..19K}
was used for the observations of a high-frequency line. The 7 pixels, spaced
by $18\arcsec$, are arranged in a hexagon around a central pixel.
The observations took place on several occasions during 2010; April 2-11, July 7,
August 4-8, and September 10-13.

\begin{figure}[!h]
\includegraphics[width=6.0cm]{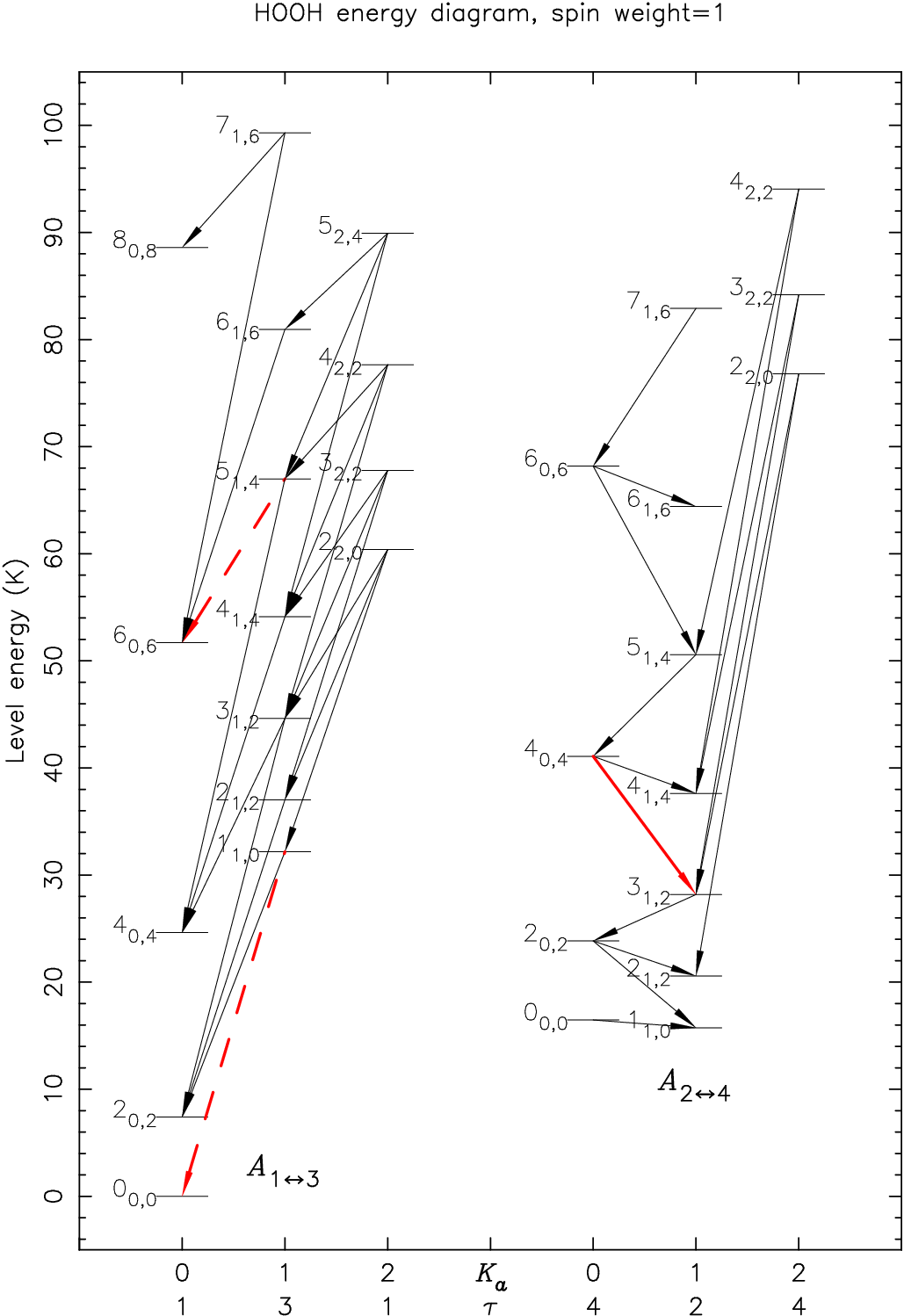}
\includegraphics[width=6.0cm]{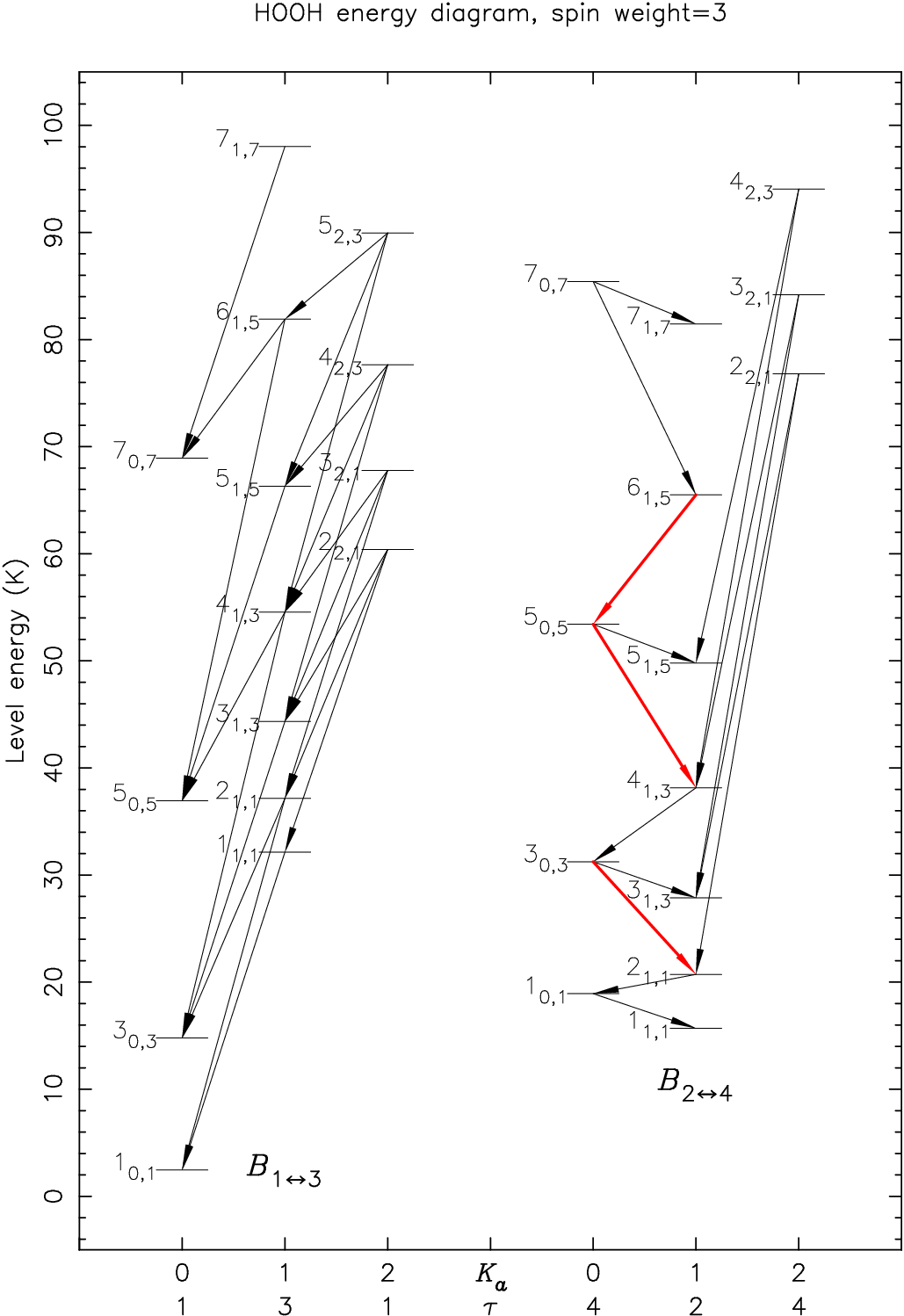}
\caption{HOOH energy level diagrams. The energy is given in K on the vertical axis,
and at the bottom the quantum numbers $K_a$ and $\tau$ are shown. To the left of
each level the rotational quantum numbers $J_{K_a,K_c}$ are listed. The upper
diagram shows the levels with a nuclear spin weight of 1 ($A_{1\lra 3}$ and
$A_{2\lra 4}$), while those levels with a spin weight of 3 ($B_{1\lra 3}$ and
$B_{2\lra 4}$) are shown in the lower diagram. The allowed $c$-type transitions
within each of the four ladders are indicated as arrows. The red arrows
represent detected lines, while the red-dashed arrows indicate non-detections.}
\label{f:egylevs}
\end{figure}

The targeted {\hooh} lines are listed in \tref{t:obslines}. The frequency, transition
quantum number designation, energy of upper level, and Einstein $A$-coefficient are
listed. All values have been compiled from the JPL catalogue
\citep{1998JQSRT..60..883P}. The frequency uncertainty for the listed lines is 0.1 MHz or
better and this corresponds to $0.14\kmps$ at 219 GHz. We also present in \fref{f:egylevs}
the energy diagrams for all levels below 100 K. Owing to the symmetry of {\hooh},
four different radiatively decoupled ladders occur: $A_{1\lra 3}$, $A_{2\lra 4}$, $B_{1\lra 3}$, and
$B_{2\lra 4}$. The subscript indicates the pair of torsional quantum numbers $\tau$
involved. The $c$-type electric dipole transitions must also obey $\tau = 1\lra 3$ or
$\tau = 2\lra 4$ \citep{1965JChPh..42.1931H} and are drawn as downward arrows
connecting the upper and lower levels in \fref{f:egylevs}. The $A$-species have a
nuclear spin weight of 1 and the $B$-species have a spin weight of 3
\citep{1984CaJPh..62.1392H}. This is, of course, due to the nuclear spin directions of
the two H atoms (in the same way as the ortho and para symmetries occur for {\hho} or
{\hhco}).

\begin{table}
\caption{Observed HOOH lines}
\label{t:obslines}
\centering
\begin{tabular}{lcccc}
\hline\hline
Frequency & \multicolumn{2}{c}{Transition} & $E_u$ & $A_{ul}$ \\
(MHz)     &  $J'_{K'_a,K'_c} - J''_{K''_a,K''_c}$ & $\tau' -\tau''$ & (K)   & (${\rm s^{-1}}$) \\
\hline
219166.86 & $3_{0,3}-2_{1,1}$ & $4-2$ & 31.2 & $8.58\times 10^{-5}$  \\
251914.68 & $6_{1,5}-5_{0,5}$ & $2-4$ & 65.5 & $2.46\times 10^{-4}$  \\
268961.17 & $4_{0,4}-3_{1,2}$ & $4-2$ & 41.1 & $1.84\times 10^{-4}$  \\
318222.52 & $5_{0,5}-4_{1,3}$ & $4-2$ & 53.4 & $3.31\times 10^{-4}$  \\
318712.10 & $5_{1,4}-6_{0,6}$ & $3-1$ & 67.0 & $4.12\times 10^{-4}$  \\
670595.82 & $1_{1,0}-0_{0,0}$ & $3-1$ & 32.2 & $5.79\times 10^{-3}$  \\
\hline
\end{tabular}
\end{table}

The ground-state symmetry species is that of $A_{1\lra 3}$ with the $B_{1\lra 3}$
state only about 2.5 K higher energy. This means that the energy difference
between the species with different nuclear spin weights is much smaller than the
corresponding difference for {\hho} or {\hhco}. The torsional $\tau = 2, 4$ states are
about 16 K above the $\tau = 1, 3$ states. This difference stems from the
tunneling through the {\it trans} barrier and is comparable to the gas kinetic
temperatures of $20 - 30\K$ found in {\roa} \citep{1990ApJ...365..269L,
2003A&A...402L..73L, 2011A&A...527A..39B}.

\begin{figure}
\includegraphics[width=6.5cm]{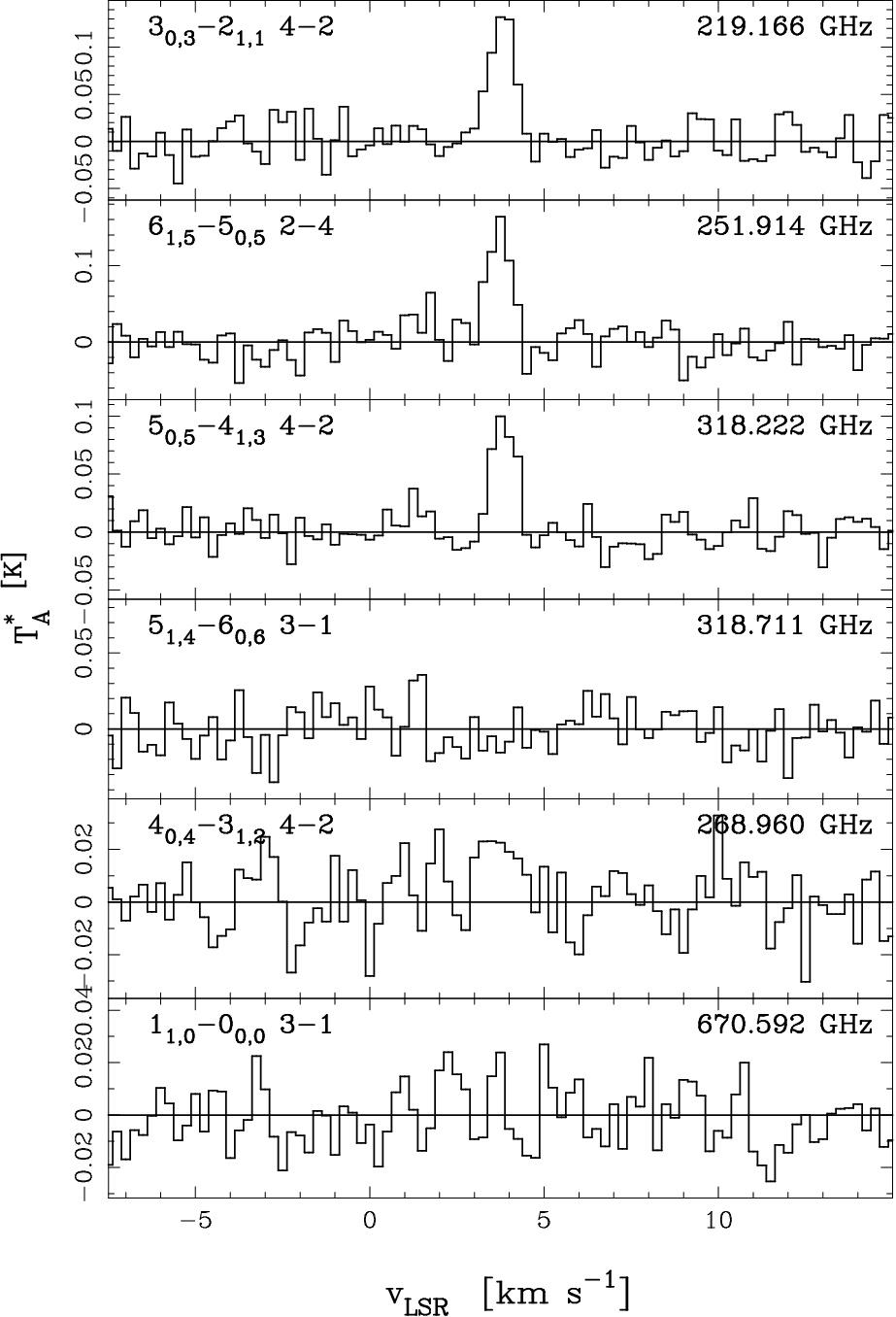}
\caption{{\hooh} spectra toward {\roa}. The transition is indicated in each spectrum.
The $T_\mathrm{A}^\ast$ intensity scale is in K and the velocity ($\vlsr$) scale
is in $\kmps$. The velocity resolution is $0.25\kmps$.}
\label{f:spectra}
\end{figure}

\begin{figure}
\includegraphics[width=6.5cm]{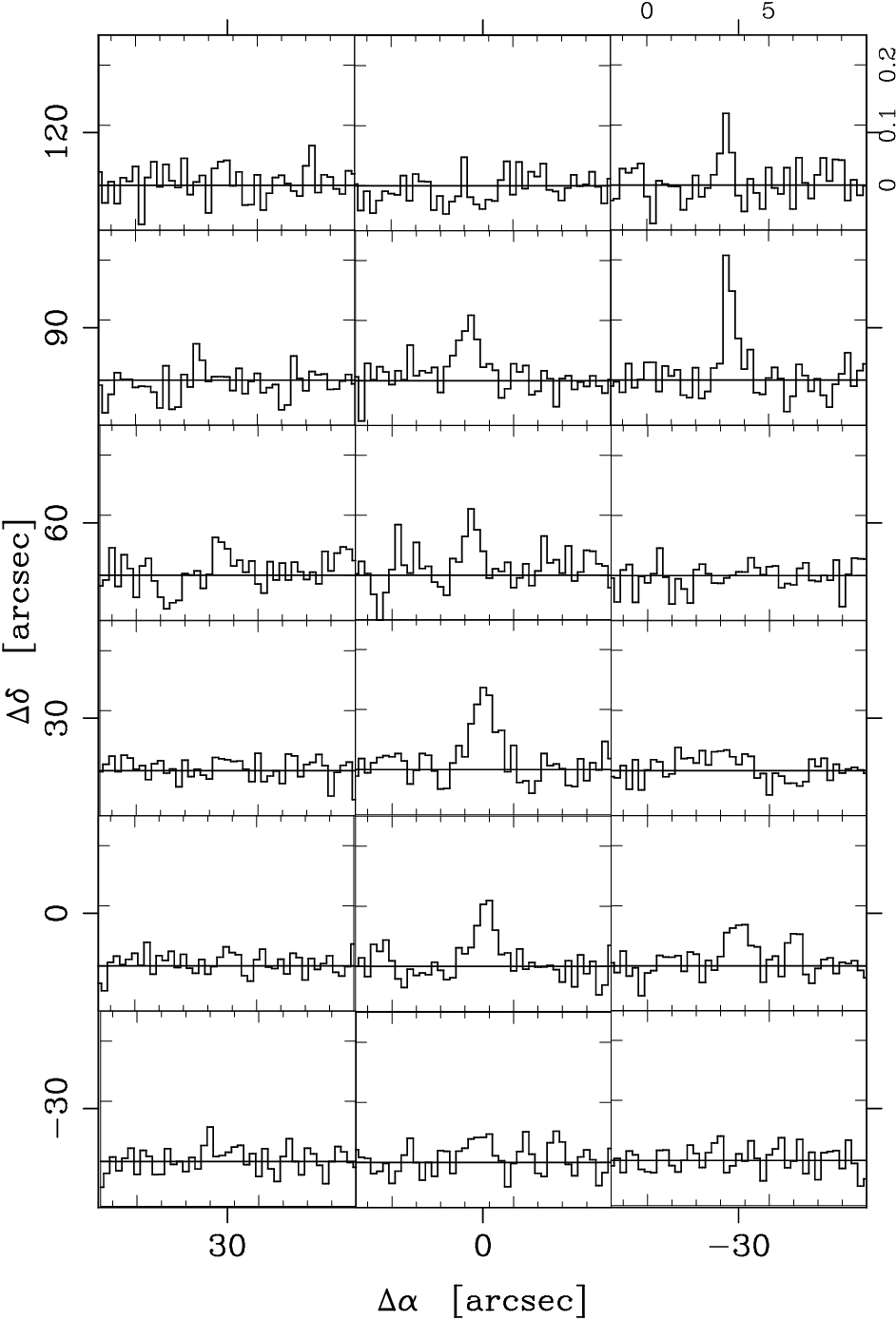}
\caption{Map spectra of the 219166 MHz {\hooh} line toward {\roa}. The map offset
$(0\arcsec, 0\arcsec)$ corresponds to the SM1 position.
The $T_\mathrm{A}^\ast$ intensity scale, in K, and the velocity ($\vlsr$) scale,
in $\kmps$, are indicated in the upper right spectrum. The velocity resolution is $0.25\kmps$.}
\label{f:map219}
\end{figure}

In \fref{f:spectra} we show the {\hooh} spectra toward the core of {\roa}. The
upper five spectra are
toward the SM1 core, \RA{J2000}{16}{26}{27.2} and \Dec{J2000}{-24}{24}{04}. The 670 GHz CHAMP$^+$
spectrum is an average of all pixels and is centered $30\arcsec$ north of the SM1 position (usually
denoted SM1N). In \fref{f:map219} spectra of the 219166 MHz line for the
observed map positions are shown. In the southern part
the line is centered at $3.7\kmps$, while further north, at offset $(0'',+60'')$,
the line is at $3.2\kmps$. This NS velocity
gradient is almost identical to the one seen for {\hhco} and its deuterated variants
\citep{2011A&A...527A..39B}. The narrow peaks to the NW are adjacent to where the sulphur species peak
as noted by the same authors. In addition to the map in \fref{f:map219}, we also mapped the
251~GHz line, albeit with a poorer S/N ratio and will not discuss it further here.

The beam sizes, velocity-integrated main-beam brightness temperatures
($\int T_\mathrm{mb}\, dv$) , fitted LSR velocities ($v_{\rm LSR}$), and line widths
(FWHM) are tabulated in \tref{t:lineints}. The $T_\mathrm{mb}$-scale was
established assuming main-beam efficiencies of 0.75, 0.73, and 0.4, for APEX-1,
2, and CHAMP$^+$, respectively. The upper limits are $3\sigma$ and for the
CHAMP$^+$ line two intensities are listed in \tref{t:lineints}; one for the
three pixels closest to the SM1 position and the other by averaging data from
all pixels. The tabulated errors and upper limits depend only on the channel
noise.

\begin{table}
\caption{Observed {\hooh} line velocities, widths, and intensities}
\label{t:lineints}
\centering
\begin{tabular}{lrrrr}
\hline\hline
Freq. & Beam size & $v_{\rm LSR}$ & FWHM & $\int T_\mathrm{mb}\, dv$ $^a$ \\
(MHz)     & (arcsec) & ($\kmps$) & ($\kmps$) & (${\rm K\, km\, s^{-1}}$) \\
\hline
219167 & 28 & 3.8 & 0.84  & 0.167(0.018)  \\
251915 & 25 & 3.7 & 0.75  & 0.165(0.018)  \\
268961 & 23 & 3.7 & (1.2) & 0.040(0.011)  \\
318223 & 20 & 3.8 & 0.78  & 0.106(0.013)  \\
318712 & 20 &     &       & $< 0.044$     \\
670596 & 9  &     &       & $< 0.12^b$    \\
       &                         &     &       & $< 0.051^c$   \\
\hline
\multicolumn{5}{l}{$^a$integr. from 2.5 to $4.5\kmps$, errors are $1\sigma$,
upper limits are $3\sigma$} \\
\multicolumn{5}{l}{$^b$average value from three pixels closest to the SM1 position} \\
\multicolumn{5}{l}{$^c$average value from all pixels, central pixel on SM1N}
\end{tabular}
\end{table}

\begin{figure}
\includegraphics[width=7.9cm]{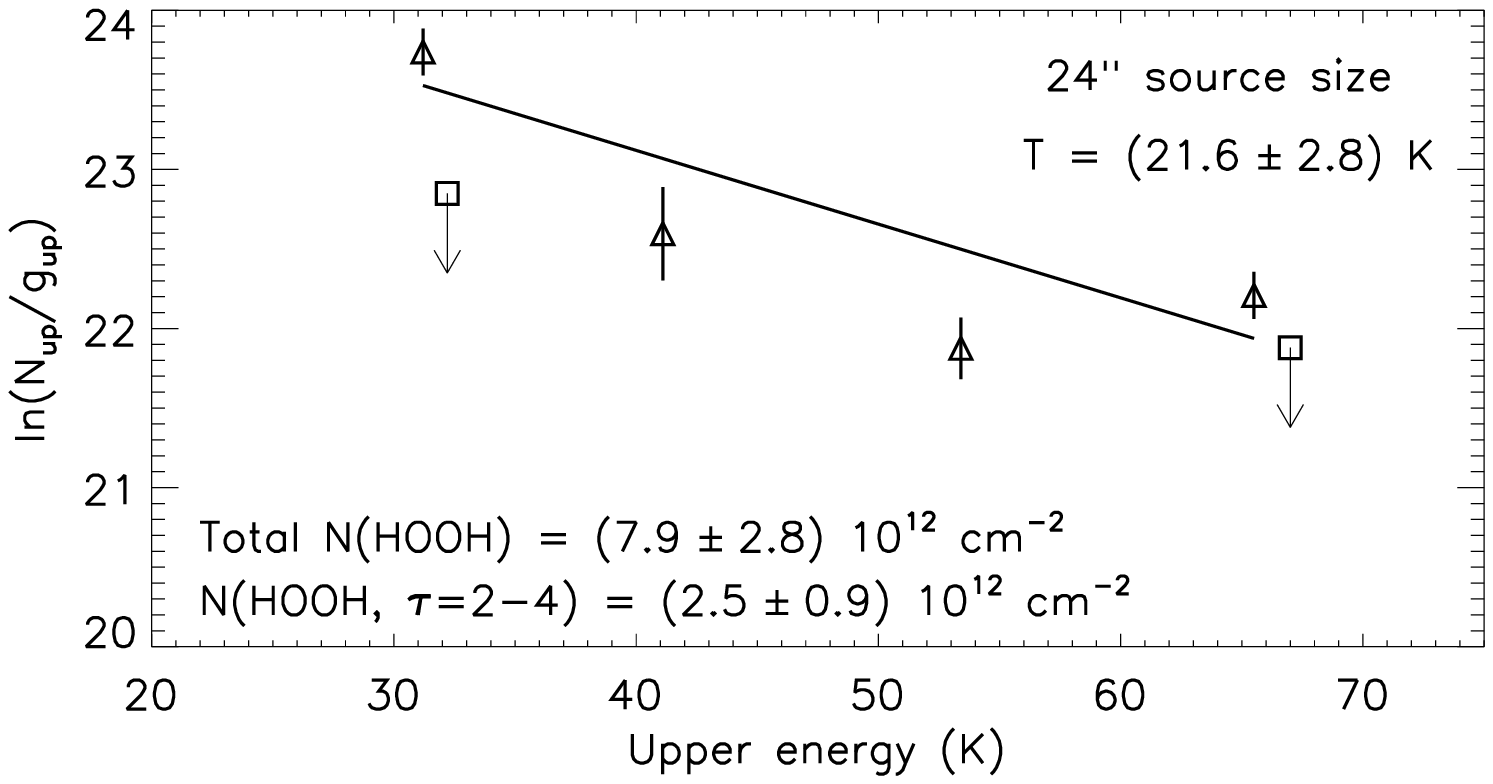}
\caption{{\hooh} rotation diagram.
The detected lines (only $\tau = 2\lra 4$ lines) are shown as open triangles with
error bars. The size of the error bars corresponds to the total uncertainty (noise
and calibration). The $3\sigma$ upper limits (not included in the fit) are shown as open
squares and downward arrows and originate from transitions between $\tau = 1, 3$
states. The fitted rotation temperature is noted, as is the
total column density. The column density for unpopulated
$\tau = 1, 3$ states is also given.}
\label{f:rotdia}
\end{figure}

Using the integrated line intensities in \tref{t:lineints} (corrected by a beam-filling
factor corresponding to a source size of $24''$, see \citet{2011A&A...527A..39B}), we
performed a rotation diagram analysis \citep{1999ApJ...517..209G}. From this we can
determine the rotation temperature, $\trot$, as well as the {\hooh} column density,
$N(\mathrm{HOOH})$. The resulting rotation diagram is displayed in \fref{f:rotdia}. The
fit is based on the detected $\tau = 2\lra 4$ lines. We derive $\trot = 22\pm 3\,\K$ and
a total {\hooh} column density of $\me{(8\pm 3)}{12}\cmcm$ where the errors depend on
the uncertainty of the integrated intensities and a calibration uncertainty of $10-15\%$.
This rotation temperature is very close to the estimates of the kinetic temperatures
from {\hhco} at the SM1 position. Obviously, the non-detection of the $\tau =
1\lra 3$ lines (open squares in \fref{f:rotdia}) is not consistent with the fit.
Especially the 670 GHz line should have been detected given these values of $\trot$ and
$N(\mathrm{HOOH})$.  Hence, we conclude that the $\tau = 1, 3$ states are not
populated according to a simple LTE model. The non-detection of the 670 GHz line
could be a result of subthermal excitation. The spontaneous rate coefficient is large
(\tref{t:obslines}) and for a typical value of a collision coefficient
($\sim 10^{-10}\cul$), the critical density of the 670 GHz transition is about two orders
of magnitude higher than the {\hh} density determined from {\hhco} and {\chhhoh}
observations in the same source \citep{2011A&A...527A..39B}. From the energy
diagrams (\fref{f:egylevs}) it is clear that there is a lack of radiative
de-excitation routes out of the $K_a=0$ levels in the $\tau= 1, 3$ states (as opposed to the
$\tau= 2, 4$ states). The population of the
$\tau= 1, 3$ states could therefore be confined to the $K_a=0$ levels if the collisional
excitation is inefficient. However, a full statistical equilibrium analysis is needed
for understanding the details of the excitation. This also requires some basic
knowledge of the collision coefficients. For now we assume
that the $\tau = 1, 3$ population is
negligible compared to the $\tau = 2, 4$ population. In this case, we
instead find a total molecular column density of $\me{(3\pm 1)}{12}\cmcm$.

\section{Discussion}
\label{s:dis}
Given the good agreement of the velocities of our detected {\hooh} lines
(together with the $\approx 0.1\kmps$ accuracy of the laboratory frequencies) with those from other
species we are very confident that the lines belong to {\hooh}. From our mapping
of the 219 GHz line it is also evident that the NS velocity gradient seen for {\hooh}
reflects that of other species. Moreover, the derived
rotation temperature of $22\pm 3$ K is what would be expected for the {\roa} cloud core.
For this fairly low excitation temperature one would not expect many lines from
other species to be present and, using the JPL and Cologne databases
\citep{1998JQSRT..60..883P, 2005JMoSt.742..215M}, we
found no lines from other species that could possibly interfere with the identification.
Of course, the narrow lines (with FWHM typically $< 1\kmps$) seen toward the {\roa}
cloud core also make line confusion much less likely.

From the {\hhco} and {\chhhoh} analysis of the SM1 core \citet{2011A&A...527A..39B} determined
an {\hh} column density of $\me{3}{22}\cmcm$. Assuming that the {\hooh}
level populations mainly reside in the $\tau = 2, 4$ states, we arrive
at an {\hooh} abundance of about $\me{1}{-10}$. This is well below
the limit of $\me{4.5}{-10}$ found toward Orion KL by \citet{1987ApJ...315..621B}.

According to current gas-phase schemes (e.g., the OSU chemical reaction 
database), formation of {\hooh} in the gas phase is not
efficient. Only two very slow reactions are proposed for its formation, 
reaction of {\hh} with {\hoo}, or reaction of two OH radicals.

On grains, {\hooh} is formed through successive hydrogen additions to 
{\oo}. This was first proposed by \citet{1982A&A...114..245T},
based on theoretical arguments.
Recent laboratory experiments have been made to investigate this route, 
up to formation of water molecules:
\begin{eqnarray}
\mathrm{O_2 + H + H} \to \mathrm{HOOH}  \\
\mathrm{HOOH + H} \to \mathrm{H_2O + OH}.    
\end{eqnarray}

\citet{2008CPL...456...27M} have investigated the reaction of H atoms with solid 
{\oo} at 10\,K. Subsequently, \citet{2008ApJ...686.1474I} have
investigated the same reaction in the temperature range 12-28\,K. Both 
studies showed that the conversion of {\oo} stops at some
point before exhaustion of {\oo} because of shielding of {\oo} in the 
deepest layers. The experiment of \citet{2008ApJ...686.1474I} shows that
the shielding decreases with increasing temperature, pointing to the 
fact that the {\oo} ice may become more porous when close
to its sublimation temperature (30\,K).

Because this shielding may not be very relevant in space,
\citet{2009ApJ...701..464O} investigated the formation of {\hooh} and {\hho} when 
codeposing O$_2$ and H in the temperature range 10-40\,K.
The {\hho}/{\hooh} ratio in the formed ices is observed to depend 
strongly on the temperature, and on the {\oo}/H flux.
The measured {\hho}/{\hooh} ratio is lower than 5 in all experiments 
($T=$ 10, 20\,K and {\oo}/H-flux between $\me{3.8}{-4}$ and
$\me{1.9}{-2}$), although higher values may be obtained in the 
case of lower {\oo}/H-flux.
The present detection of {\hooh} may open the possibility to quantify the 
importance of reactions (1) and (2) in the formation of water.

\citet{2008ApJ...686.1474I} have modeled the formation of water in typical dense 
clouds (their Figure 4). The laboratory results
lead to a revision of the energy barriers involved in the models, and 
their new model (only accounting for the three main
routes of water formation on the grains) predicts a fractional abundance 
for {\hooh} of a few $10^{-14}$ with respect to H nuclei.
This is more than three orders of magnitude lower than our detection.

Further understanding will require detailed chemical modeling of grain 
chemistry. This will be the scope of a forthcoming paper.
The observation of water in the {\roa} region with the {\it Herschel} 
Observatory as well as the confirmation of the {\oo} detection
would also be very valuable in setting constraints on the models.

\begin{acknowledgements}
  We acknowledge the excellent observational support from the APEX staff. We are
  grateful to A. Gusdorf for doing some of the CHAMP$^+$ observations. BP is
  funded by the Deutsche Forschungsgemeinschaft (DFG) under the Emmy Noether
  project number PA1692/1-1.
\end{acknowledgements}
\bibliographystyle{aa} 
\bibliography{myrefs} 
\end{document}